\documentclass[twocolumn]{aastex62}
\UseRawInputEncoding
\usepackage{verbatim}
\usepackage{gensymb}
\usepackage{graphicx}
\usepackage{lineno}
\usepackage{array}
\usepackage{booktabs}
\usepackage{natbib}
\usepackage{bm}
\usepackage{amsmath}
\newcommand{\otoprule}{\midrule[\heavyrulewidth]}
\newcolumntype{+}{>{\global\let\currentrowstyle\relax}}
\newcolumntype{^}{>{\currentrowstyle}}

\begin{document}

\title{\textbf{First detection of HC$_5$N in a Class II disk around TW Hya}}

\correspondingauthor{Steven C. Wampler}
\email{swampler@nrao.edu}

\author{Steven C. Wampler}	
\affiliation{National Radio Astronomy Observatory, Charlottesville, VA 22903, USA}

\author{Ryan A. Loomis}
\affiliation{National Radio Astronomy Observatory, Charlottesville, VA 22903, USA}

\author{Amina Diop}
\affiliation{Department of Astronomy, University of Virginia, Charlottesville, VA 22904, USA}

\author{L. Ilsedore Cleeves}
\affiliation{Department of Astronomy, University of Virginia, Charlottesville, VA 22904, USA}

\author{Yuri Aikawa}
\affiliation{Department of Astronomy, Graduate School of Science, The University of Tokyo, Tokyo 113-0033, Japan}

\author{Romane Le Gal}
\affiliation{Institut de Plan\'{e}tologie et d’Astrophysique de Grenoble, FR}
\affiliation{Institut de Radioastronomie Millim\'{e}trique,Universit\'{e} Grenoble Alpes, Saint-Martin-d’H\'{e}res, Auvergne Rh\^{o}ne-Alpes, FR}

\author{Viviana Guzman}
\affiliation{Institute of Astrophysics, Pontificia Universidad Cat\'{o}lica de Chile, Vicu\~{n}a Mackenna 4860, Macul, Regi\'{o}n Metropolitana, Santiago, Chile}
\affiliation{$^{}$Millennium Nucleus on Young Exoplanets and their Moons (YEMS)}

\author{Charles J.\ Law}
\altaffiliation{NASA Hubble Fellowship Program Sagan Fellow}
\affiliation{Department of Astronomy, University of Virginia, Charlottesville, VA 22904, USA}

\author{Karin \"Oberg}
\affiliation{Center for Astrophysics, Harvard \& Smithsonian, 60 Garden St., Cambridge, MA 02138, USA}

\author{Jamila Pegues}
\affiliation{Department of Astronomy, University of Virginia, Charlottesville, VA 22904, USA}

\author{Richard Teague}
\affiliation{Department of Earth, Atmospheric, and Planetary Sciences, Massachusetts Institute of Technology, Cambridge, MA 02139, USA}

\author{Catherine Walsh}
\affiliation{School of Physics and Astronomy, University of Leeds, Leeds LS2 9JT, UK}

\begin{abstract}
Over the last decade of ALMA's operation the molecular inventory of protoplanetary disks has expanded rapidly, revealing a diverse set of nitrogen bearing organics and carbon chain molecules that trace both prebiotic chemistry and fundamental disk physics. Despite this progress, detections of larger species such as cyanopolyynes have remained limited, leaving larger carbon-chain chemistry in Class II disks largely unconstrained. Here, we report the first detection of HC$_5$N toward the TW Hya protoplanetary disk, representing the largest cyanopolyyne identified to date in a Class II system.  We derive an HC$_5$N column density for two rotational transitions J=41-40 and J=37-36 $N_T \sim10^{12}$ cm$^{-2}$ for assumed $T_{rot}$~=~20-50 K and optically thin emission in LTE. We compare HC$_5$N and HC$_3$N formation mechanisms and analyze the HC$_3$N / HC$_5$N ratio. We use a chemical model to estimate the expected abundance and emitting layer of HC$_5$N in a TW~Hya--like disk. Although HC$_5$N emission is spatially unresolved and measured column densities suggest an origin in the warm molecular layer where CN based pathways are active. This detection extends the known carbon-chain chemistry in Class II disks and demonstrates that long cyanopolyynes can form and persist in planet forming environments.
\end{abstract}

\section{Introduction}

Protoplanetary disks contain the molecular reservoirs that ultimately seed the formation of planets, moons, and comets. An increasing number of confirmed exoplanets is revealing a wide diversity of physical and chemical environments \citep[e.g.,][]{Akeson_2013}, and with the advent of JWST, exoplanet atmospheres are now being characterized directly. Linking these atmospheric measurements to the chemical initial conditions of planet formation requires a detailed understanding of the molecular inventories present in disks. With ALMA observations offering the opportunity to connect the disk chemistry of evolving stellar systems to the chemical inventories of mature planets. Understanding how disk chemistry evolves requires sampling systems across stellar masses and evolutionary stages, since molecular abundances and physical conditions vary substantially between young Class I disks, older Class II disks, and disks around T Tauri and Herbig Ae stars.

Complex organic molecules (COMs) are readily observed in the warm, dense regions surrounding Class 0 protostars \citep[e.g.,][]{Jorgensen_2016}. Recent work has shown that COMs can accrete onto disks from the surrounding envelope and are present during the early stages of star formation \citep{Murillo_2022}. COMs have also been detected in a small number of warmer, high mass Herbig Ae disks \citep[e.g.,][]{Booth_2024, Booth_2025a, Yamato_2024, Booth_2025b}. By contrast, detections of COMs in Class II T Tauri disks are exceedingly rare with only two species CH$_3$OH \citep{Walsh_2016} and CH$_3$CN \citep{Oberg_2015} detected to date. This scarcity most likely reflects the extremely low gas-phase abundances of these COMs in disks, where they are readily destroyed, frozen out, or inefficiently produced \citep[e.g.,][]{Walsh_2014}. What remains unclear then is the link between the chemically-rich environments of young protostellar systems and their older and colder counterparts. The rarity of COM detections in T Tauri disks limits our ability to assess how much chemical complexity is inherited versus produced in situ, while observations of simpler species suggest that disk physical conditions, including variations in the C/O ratio, play a major role in regulating this chemistry \cite[e.g.,][]{Oberg_2011, Oberg_2015, Bergner_2018, Oberg_2023, Aikawa_Okuzumi_Pontoppidan_2024}. 

While CH$_3$CN is the only nitrogen-bearing COM detected in Class II T Tauri disks to date, other smaller nitrile species, characterized by the presence of a cyano group (-CN), have emerged as critical probes for understanding the chemistry relevant to planet and comet formation. Measurements from the Rosetta mission revealed that nitriles are abundant in solar system comets \citep[e.g.,][]{Le_Roy_2015}, suggesting that they are critical to understanding the chemistry of the early solar nebula. Although large and complex nitrile species will likely undergo destruction during planet formation and on early Hadean-like exoplanet surfaces, the cyanide group itself can persist under these conditions \citep{Zahnle_1986, Powner_2009}. Beyond their importance to pre-biotic chemistry, the cyano functional group imparts a large dipole moment and hyperfine splitting, yielding multiple strong rotational transitions that are accessible within a single observation. These unique properties make them effective tracers of disk structure constraining elemental abundances, isotopic ratios both radially and vertically within disks, turbulence, and constraining magnetic fields. \citep[e.g.,][]{Guzman_2017, Bergner_2021, Cataldi_2021, Teague_2016, Paneque-Carreno_2024, Teague_2025} 

Cyanopolyynes, which are long unsaturated carbon chain nitriles of the form HC$_n$N where n = 3,5,7\ldots, are frequently detected in other interstellar environments including the envelopes of evolved stars and cold dark clouds \citep[e.g.,][]{Broten_1978}, and their chemistry is largely understood to proceed via gas phase reactions \citep[e.g.,][]{Fukuzawa_1998, Taniguchi_2017}. Their precursors CN, C$_2$H$_2$, and HCN have long been known to exist in disks \citep[e.g.,][]{Dutrey_1997, van_Zadelhoff_2001, Thi_2004, Gibb_2007, Carr_2008}, but detections of larger HC$_n$N species have only been made in other environments, extending up to HC$_{11}$N \citep[e.g.,][]{Loomis_2021}. To date, HC$_3$N remains the only cyanopolyyne detected in a Class II disk \citep[e.g.,][]{Chapillon_2012, Oberg_2015, Bergner_2018, Ilee_2021, Phoung_2021}. In the interstellar medium, cyanopolyynes serve as valuable tracers of both physical and chemical conditions, providing diagnostics of temperature, turbulence, and isotopic fractionation \citep[e.g.,][]{Taniguchi_2019b}. Their utility for isotopic studies arises from their carbon chain structure, each molecule contains both multiple carbon atoms and a cyano functional group, allowing for constraints on carbon and nitrogen fractionation \citep[e.g.,][]{Sakai_2011a, Taniguchi_2017, Burkhardt_2018}. Cyanopolyynes also have large dipole moments that facilitate reliable determinations of column densities ($N_T$) and rotational temperatures ($T_{\rm rot}$) \citep[e.g.,][]{Kastner_1997, Dutrey_1997, Chapillon_2012, Bergner_2018}, and as the carbon chain length increases, the rotational transitions become more closely spaced in frequency, enabling more efficient coverage within a single spectral setup for bandwidth-limited observations. These factors make detections of longer cyanopolyynes in disks particularly compelling to pursue.

These considerations motivate the search for longer cyanopolyynes in disks. The protoplanetary disk around TW Hya, located at a distance of ~60 pc \citep{Gaia_Collaboration}, provides an exceptional laboratory for such a study. As the nearest T Tauri star, with a stellar mass of 0.88 M$_\odot$ and spectral type K7, TW Hya is often regarded as a close analog to the early solar nebula \citep[e.g.,][]{Bergin_2013}. Its bright, nearly face-on disk has been the subject of extensive observational campaigns, yielding detailed constraints on both its dust distribution \citep[e.g.,][]{Andrews_2016, Huang_2018} and chemistry \citep[e.g.,][]{Walsh_2016, Favre_2018, Loomis_2018a, Hily-Blant_2019, Oberg_2021, Terwisscha_van_Scheltinga_2021, Cleeves_2021, Calahan_2021}.

In this paper, we present the first detection of HC$_5$N toward TW Hya. Our observation setup and data reduction are discussed in Section 2. The detection results and a column density analysis are presented in Section 3. In Section 4, we discuss the morphology and formation chemistry of HC$_5$N. Finally, we summarize our results and implications and outline prospects for future observations.

\section{Observations}
    \subsection{Observational Details}
          TW Hya was observed over 7 execution blocks on 22-Mar-2021, 23-Mar-2021, 25-Mar-2021, and 28-Mar-2021 in Band 3 as part of the ALMA Cycle 6 project 2019.1.00769.S. The total on-source integration time was 325 minutes. The correlator setup was identical for each execution block and included a Time Division Mode (TDM) continuum window centered at 110.169 GHz with a bandwidth of ~2GHz as well as a Frequency Division Mode (FDM) spectral windows centered at 99.665, 100.078, 98.541, 96.982, 109.173, and 108.715~GHz. These spectral windows had a bandwidth of 58.594 MHz and channel spacing of 30.518 kHz ($\sim$0.08~km s$^{-1}$). The main targets of the observations were HC$_3$N and HC$_3$N isotopologues, but two transitions of HC$_{5}$N were targeted as ancillary bonus lines (see Table \ref{table_1}).

        \begin{deluxetable*}{ccccccc}
        \caption{Observed Transition Properties}
        \label{table_1}
        \tablecolumns{7}
        \tablewidth{\textwidth}
        \tablehead{
        \colhead{Species} &
        \colhead{Transition} &
        \colhead{Frequency\tablenotemark{a}} &
        \colhead{\ensuremath{E_u}} &
        \colhead{\ensuremath{S_{ij}\mu^2}\tablenotemark{b}} &
        \colhead{Int.\ Flux Dens.\tablenotemark{c}} &
        \colhead{Filter Response\tablenotemark{d}} \\
        & &
        \colhead{(MHz)} &
        \colhead{(K)} &
        \colhead{\ensuremath{(\mathrm{D}^2)}} &
        \colhead{\ensuremath{(\mathrm{mJy\,km\,s^{-1}})}} &
        \colhead{\ensuremath{(\sigma)}}
        }
        \startdata
        \multicolumn{7}{l}{HC$_3$N} \\
        -- & J=12--11 & 109173.63 & 34.05 & 167.09 & 275$\pm$28 & -- \\
        \multicolumn{7}{l}{HC$_5$N} \\
        -- & J=37--36 & 98512.524 & 85.10 & 2080.91 & 3.7$\pm$3.5 & 3.4$\sigma$ \\
        -- & J=41--40 & 109160.973 & 104.78 & 2305.96 & 8.3$\pm$6.3 & 4.2$\sigma$
        \enddata
        \tablenotetext{a}{Center frequency of collapsed hyperfine components (spacing smaller than channel width).}
        \tablenotetext{b}{\ensuremath{S_{ij}\mu^{2}} of combined hyperfine components.}
        \tablenotetext{c}{Velocity-integrated between 2.0--3.7~km~s$^{-1}$.}
        \tablenotetext{d}{The HC$_3$N emission was used as the filter; therefore a detection significance is not meaningful.}
        \end{deluxetable*}

    \subsection{Calibration and Imaging}
        The data were initially calibrated by the ALMA Observatory using the ALMA pipeline version 2020.1.0.40 \citep{Hunter_2023} in CASA 6.1.1.15. The quasar J1037-2934 was used for phase calibration and J1107-4449 was used as the bandpass and flux calibrator across all executions. We then additionally self-calibrated the data with three rounds of phase-only and one round of phase and amplitude self-calibration.

        For spectral line imaging, we continuum subtracted the self-calibrated visibilities in the \textit{uv}-plane and \texttt{CLEAN}ed the lines using natural weighting for maximum sensitivity in CASA 6.1.1.15. This resulted in a beam of 1.07" x 0.86" $-84.531^\circ$ for the HC$_5$N J=37-36 transition and 0.97" x 0.77" $-85.550 ^{\circ}$ for the HC$_5$N J=41-40 and HC$_3$N J=12-11 transitions, due to their different rest frequencies. These correspond to roughly $\sim$60 AU resolution at TW Hya's distance. For HC$_3$N a Keplerian mask based on the adopted TW~Hya disk geometry: $i=5^\circ$, ${\rm PA}=152^\circ$ and system parameters $M_{\ast}=0.88\,M_\odot$, $d=59.5$ pc, with a radial sampling of $dr=0.1$ \citep{Huang_2018,Andrews_2012,Gaia_Collaboration} was generated for \texttt{CLEAN}ing, while no mask was used for the HC$_5$N transitions as there was no significant emission to \texttt{CLEAN}. The achieved rms values were $\sim$1.1 and $\sim$1.3 mJy bm$^{-1}$ respectively for the J=37-36 and J=41-40 HC$_5$N transitions, and $\sim$1.3 mJy bm$^{-1}$ for HC$_3$N J=12-11.

\section{Results}

    \subsection{HC$_3$N and HC$_5$N Integrated Intensity}
\begin{figure*}
        {\includegraphics[width=\textwidth]{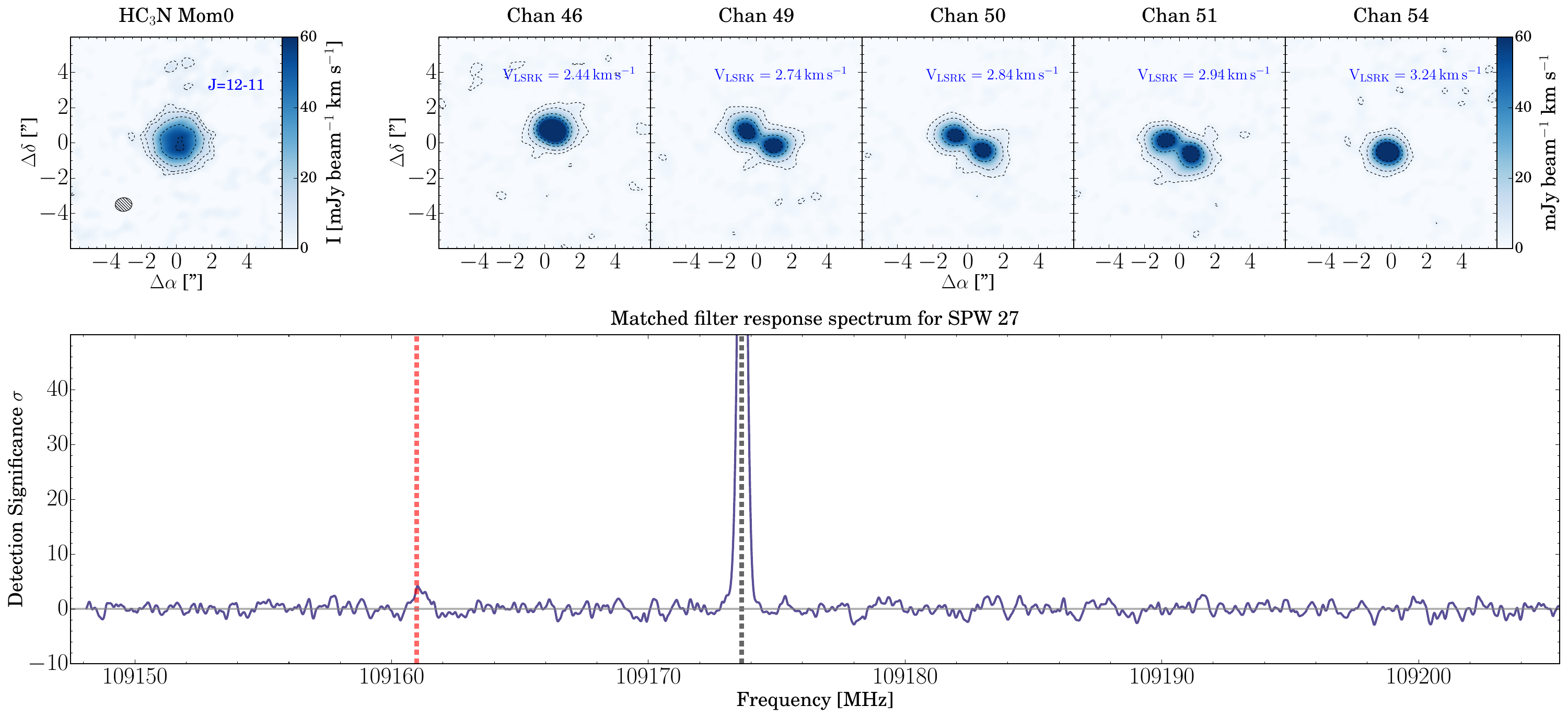}}
        \caption{HC$_3$N $J=12$-$11$} emission. Upper left panel: Moment 0 map of HC$_3$N emission. Top center panels: Channel maps of the HC$_3$N line; selected velocity channels are shown to highlight the Keplerian rotation of the disk. Bottom panel: Matched filter response spectrum for the full spectral window containing the HC$_3$N $J=12$-$11$ transition. The red dashed line marks the expected rest frequency of HC$_5$N $J=41$-$40$.
        \label{figure_1}
    \end{figure*}

    HC$_3$N provides a useful reference for our analysis, as it is the only cyanopolyyne previously detected in T Tauri disks \citep[e.g.,][]{Chapillon_2012, Bergner_2018, Ilee_2021, Calahan_2023} and was the primary target of the observations. Figure \ref{figure_1} (top left) shows a moment 0 map displaying the total integrated intensity of the strongly detected J=12-11 HC$_3$N transition. Channel maps (top row) of several select channels show the typical characteristics of Keplerian emission and the overall morphology of the HC$_3$N within the disk. 
    
\begin{figure*}
        \centering
        \includegraphics[width=0.65\textwidth]{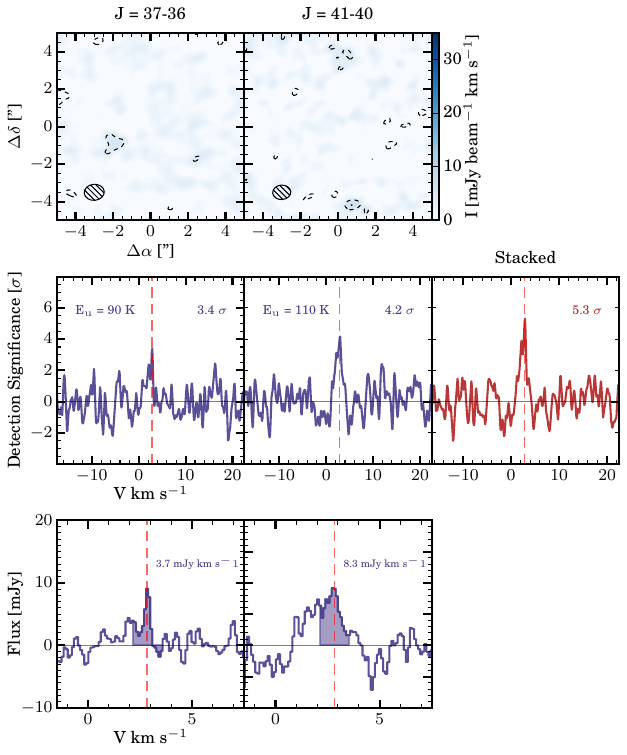}
        \caption{HC$_5$N emission detected using multiple analysis techniques. Top panels: Moment-0 maps of both observed HC$_5$N transitions. Center panels: Matched filtering impulse response spectra of the individual transitions with detection significance labeled (blue), along with the stacked combined spectrum (red). Bottom panels: Shifted and stacked image plane spectra for each transition, together with the flux integrated between 2.0--3.7~km~s$^{-1}$.}
        \label{figure_2}
    \end{figure*}

    Figure \ref{figure_2} (top row) shows the subsequent integrated intensity maps of the two targeted HC$_5$N lines. These moment maps show no strong signal.

    \subsection{Matched Filtering Analysis}
    Given the lack of an obvious image plane detection of the individual HC$_5$N transitions, we employed a matched filtering technique \citep[e.g.,][]{Loomis_2018b} to enhance line detectability. This approach exploits the fact that HC$_5$N is expected to trace roughly the same spatio-kinematic structure as other species within the disk. We examined multiple filter kernels, all of which had similar results, and thus use here an empirical filter kernel from the strong HC$_3$N J=12-11 emission illustrated in Figure \ref{figure_1}, with the assumption that HC$_3$N and HC$_5$N likely share a similar morphology. This kernel was applied to the observed visibilities across all spectral windows.
    
    Figure \ref{figure_1} (bottom panel) shows the spectral response function for the full spectral window 27, with an expected strong response for the HC$_3$N J=12-11 transition. The figure shows the expected rest frequency of the targeted HC$_5$N J=41-40 transition indicated by a dashed red line. This analysis yielded moderately significant responses at the targeted HC$_5$N frequencies, with individual line significances of 3.4$\sigma$ and 4.2$\sigma$ for the J=37-36 and J=41-40 transitions, respectively (Figure \ref{figure_2}, middle row). A stacked spectrum of the two transitions produces a 5.3$\sigma$ detection of the species.

    \subsection{GoFish Spectra and Flux Measurements}
    Having identified these potential weak HC$_5$N transitions, we next measured their integrated fluxes. We applied the \texttt{GoFish} package \citep{Teague_2019} to perform a shift and stack analysis \citep{Yen_2016}, correcting for Keplerian rotation across the disk. The well established parameters of TW Hya were adopted: inclination $i = 5^\circ$, position angle $152^\circ$, distance 60 pc, and stellar mass 0.88 M$_\odot$ \citep[e.g.,][]{Huang_2018}\citep{Gaia_Collaboration}. Spectra were extracted within a radius of 90 AU, based on the emission extent of HC$_3$N (Figure \ref{figure_1}).

    Unlike the matched-filter analysis shown in Figure 2 (middle row), which reports signal-to-noise in units of significance, the GoFish spectra are expressed in flux density (mJy). 

    GoFish analysis of HC3N J=12-11 shows emission from 2.0--3.7~km~s$^{-1}$  which is typical of keplerian emission around TW Hya \citep{Huang_2018}. Using the same integration range between 2.0--3.7~km~s$^{-1}$ yields fluxes of 3.7~$\pm$~3.5~mJy~km~s$^{-1}$ and 8.3~$\pm$~6.3~mJy~km~s$^{-1}$ for the two HC$_5$N transitions. This provides the best possible measure of the total flux, however due to low signal to noise, this does not alone confirm the detection of these transitions.  
        
    \subsection{Column Density Analysis}

    These measured fluxes are quite uncertain, but allow us to roughly constrain a disk averaged column density for different assumed excitation temperatures of HC$_5N$ in TW Hya. Rather than attempting to construct a rotation diagram from two highly uncertain flux values, we opt to directly calculate column densities from the two fluxes and some assumed rotational temperature, and then average them. Following \cite{Loomis_2018a}: assuming optically thin emission, we have: 
    
    \begin{equation} 
    N_T = \frac{4\pi S_\nu\Delta\nu Q(T_{rot})}{A_{ul} \Omega h c g_u e^{-E_u/kT_{rot}}}, 
    \end{equation}

    Where $S_\nu \Delta\nu$ is the integrated line flux, $\Omega$ is the solid angle subtended by the emitting region, $A_{ul}$ is the Einstein A coefficient, $\Delta\nu$ is the line FWHM, $E_u$ and $g_u$ are the upper-state energy and degeneracy, and $Q(T_{\rm rot})$ is the molecular partition function evaluated at the assumed rotational temperature $T_{\rm rot}$.

    Using our measured flux estimates from the GoFish analysis, we can then estimate column densities of HC$_5$N for a range of ~20K-50K that are reasonable rotational temperatures for Class II T Tauri disks \citep[e.g.,][]{Aikawa_2002, Oberg&Bergin_2021}. The average column density is presented as a range in Table \ref{table2}, propagating the range of uncertainties from the average measured fluxes of both HC$_5$N transitions.

    \begin{table}[ht!]
    \centering
    \caption{Derived HC$_5$N Column Densities}
    \label{table2}
    \vspace{-5pt}
    \begin{tabular}{cccc}
    \toprule
    $T_{\mathrm{rot}}$ [K] &
    HC$_3$N $N_T$ [cm$^{-2}$] &
    HC$_5$N $N_T$ [cm$^{-2}$] &
    Ratio \\
    \otoprule
    20 & $2.1\pm0.2\times10^{13}$ & $1.3\pm1.0\times10^{13}$ & 0.9--7 \\
    30 & $1.8\pm0.2\times10^{13}$ & $3.4\pm2.7\times10^{12}$ & 3--26 \\
    40 & $1.8\pm0.2\times10^{13}$ & $1.9\pm1.5\times10^{12}$ & 5--45 \\
    50 & $1.9\pm0.2\times10^{13}$ & $1.4\pm1.1\times10^{12}$ & 8--63 \\
    \bottomrule
    \end{tabular}
    \end{table}

    We infer an HC$_5$N column density of  $N_T \sim10^{12}$ cm$^{-2}$ for assumed $T_{\rm rot}$~=~20-50~K. These values are one to two orders of magnitude lower than typical HC$_3$N column densities reported for disks \citep[e.g.,][]{Bergner_2018, Ilee_2021}, which is reasonable when comparing to other astronomical regions where cyanopolyynes are detected \citep[e.g.,][]{Loomis_2016}. A comparison to predictions of 2D time-dependent chemical models is presented in Section 4.2.

\section{Discussion}
    In this section we examine the formation pathways and expected morphology of HC$_5$N in TW Hya and compare to a small set of chemical models to further interpret these processes.

\subsection{Formation Chemistry $\&$ Emission Region of HC$_5$N}
    Cyanopolyynes such as HC$_5$N have long been recognized as key tracers of carbon chain chemistry, with their formation pathways and physical conditions most clearly established in cold, molecule rich regions of the interstellar medium. Interpreting the presence of HC$_5$N in a protoplanetary disk therefore benefits from first considering this broader astrochemical framework. HC$_5$N was first detected in the ISM toward Sgr B2 \citep{Avery_1976} and subsequently identified in cold dark clouds in Taurus, most notably TMC-1 \citep{Little_1977}. These clouds served as ideal early laboratories for carbon chain chemistry because their low temperatures and strong UV shielding allow larger cyanopolyyne species to form efficiently and persist for long timescales. Given these conditions, the dominant formation pathway for cyanopolyynes HC$_n$N in cold clouds proceeds through gas phase neutral-neutral reactions of the general form \citep{Fukuzawa_1998}:
    \begin{equation}
    \mathrm{CN + C_{n-1}H_2 \rightarrow HC_nN + H}.
    \end{equation}
    
    The specific formation pathway for HC$_5$N is then given by \cite{Loison_2014}:
    \begin{equation}
    \mathrm{CN + C_4H_2 \rightarrow HC_5N + H}.
    \end{equation}

    Ion-molecule reactions can also contribute to carbon--chain growth in cold, highly ionized gas regions dominated by cosmic rays. In this pathway, hydrocarbon ions react with acetylene to lengthen the carbon backbone, e.g.,
    \begin{equation}
        \mathrm{C_{n}H_3^+ + C_2H_2 \rightarrow C_{n+2}H_5^+},
    \end{equation}
    after which dissociative recombination produces a longer neutral polyyne,
    \begin{equation}
        \mathrm{C_{n+2}H_5^+ + e^- \rightarrow C_{n+2}H_2 + H_2 + H}.
    \end{equation}
    These neutral chains can then react with CN to form the corresponding cyanopolyyne via the reaction shown in equation 2 providing a secondary formation channel for HC$_n$N species \citep[e.g.,][]{Herbst_1989, Fukuzawa_1998, Loison_2014}. Any carbon chain growth on grains is quickly quenched by hydrogenation reactions, making grain surface pathways an inefficient source of HC$_5$N in cold clouds.

    Physical conditions and chemical environments in protoplanetary disks differ substantially from those in cold dark clouds. Temperatures and densities throughout the structure of disks are much higher relative to cold dark clouds, and this simultaneously increases formation rates while promoting adsorption onto grains that hinder survival for carbon chains \citep{Fukuzawa_1998}. Other destruction pathways arise in disks through their X-ray induced UV radiation fields that impact the upper layers of the disk, whereas cold dark clouds are shielded from these effects \citep[e.g.,][]{Bergin_2007}. This leads to rapid photo-dissociation of carbon chains in unprotected regions. These unique disk properties additionally affect the abundance of CN \citep[e.g.,][]{Kastner_2014, Teauge&Loomis_2020}, which is the precursor to these cyanopolyynes.  These X-ray fields as well as the presence of radionuclides give rise to additional mechanisms of ionization which alter the balance of ion-molecule formation pathways in disks.\citep[e.g.,][]{Cleeves_2013} 
    
    Using the HC$_5$N/HC$_3$N column density ratio derived from our analysis in Table \ref{table2}. Across the assumed range of rotational temperatures, we find that HC$_5$N is consistently less abundant than HC$_3$N, with ratios indicating a steep decline in column density with increasing carbon-chain length. This behavior is consistent with that of large carbon chains observed in cold dark cloud environments such as TMC-1 \citep{Remijan_2006}. The similarity in relative abundances suggests that, while disk environments differ markedly in physical conditions from dark clouds, the underlying chemical trend persists in disks. The formation and survival of cyanopolyynes (HC$_n$N) becomes progressively less efficient as n increases. Our derived ratios therefore support the interpretation that disk chemistry can access long carbon-chain molecules, but with abundances that decline rapidly with molecular size, in line with expectations from both chemical kinetics and observations in other astrophysical environments. 

    \begin{figure*}[htbp]
        \centering
        \includegraphics[width=\textwidth]{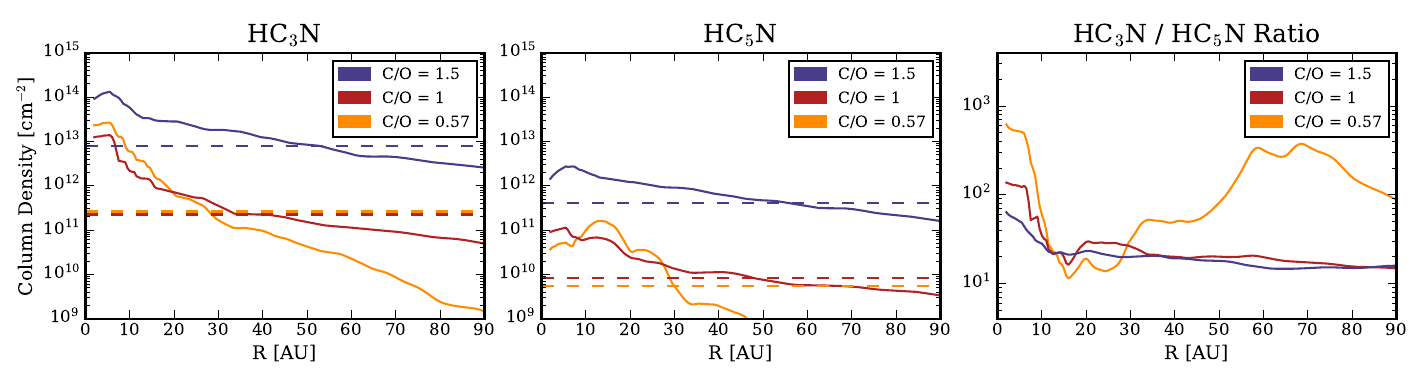}
        \caption{Comparison of HC3N (left panel) and HC5N (middle panel) column densities from chemical models assuming different C/O ratios, with dashed lines indicating disk-averaged column densities, and the HC3N/HC5N ratio (right panel) for each C/O ratio..}
        \label{figure_3}
    \end{figure*}

\subsection{ Comparison of Chemical Models and Observations}
    To aid interpretation of the observations we ran a small set of disk chemistry models. The two dimensional time-dependent chemical code used is based on the one from \cite{Fogel_2011}. The original Ohio State University (OSU) gas-phase network \citep{Smith_2004} was substantially expanded in \cite{Fogel_2011} and subsequently further updated to address additional grain surface chemistry, updated photodissociation cross-sections, and self-shielding of N$_2$ \citep{Cleeves_2014, Cleeves_2015, Cleeves_2018} as well as high temperature gas-phase reactions \citep{Anderson_2021} and updated sulfur chemistry \citep{Williams_2025}. Additional updates to several binding energies were made and will be described in a forthcoming paper.
    
    The disk physical structure is described using a standard self-similar viscous surface-density prescription, with an exponential taper beyond a characteristic radius R$_c$ of 40~AU, and a total disk mass of 0.1~$M_\sun$. The high energy stellar FUV and X-ray radiation field within the disk is calculated using the Monte Carlo code and cross sections from \cite{Bethell_2011a}, see \cite{Cleeves_2013, Cleeves_2014} for details. The TW Hya FUV spectrum from \cite{Herczeg_2002, Herczeg_2004} and the X-ray spectrum from \cite{Cleeves_2013} are used. We additionally assume a cosmic ray ionization rate of $\zeta_{\rm CR} = (1.3)\times10^{-18}$ s$^{-1}$ \citep{Cleeves_2015}. A 90$\%$ fraction of large dust grains is assumed, and the model is evolved for 1~Myr.
    
    Three different C/O ratios were considered: 0.57 (solar-like), 1.0, and 1.5. TW Hya shows significant evidence for enhanced C/O ratios in its outer disk \citep{Bergin_2016, Kama_2016, Cleeves_2021}, so we expect the elevated C/O model to more accurately describe our observations. Radial profiles of vertically integrated column densities of HC$_3$N (left panel) and HC$_5$N (right panel) from the models are shown in Figure \ref{figure_3}. Both show monotonically decreasing column densities with radius and elevated column densities when the C/O ratio is enhanced beyond 1.0, as expected. The HC$_3$N disk-averaged column density is roughly a few times 10$^{13}$~cm$^{-2}$, in line with our observations, while the HC$_5$N disk-averaged column density is $\sim$10$^{12}$~cm$^{-2}$. This value is in moderate agreement with our estimated column density, especially with a higher assumed rotational temperature of $\geq$30K, as might be expected for formation in the higher disk layers. For both HC$_3$N and HC$_5$N, the inner 10~AU shows enhanced formation, but we caution that interpretation of these regions of the disk may be difficult both due to model resolution as well as proposed C/O depletion in the inner TW Hya disk \citep{Bosman_and_Banzatti_2019}. More realistic modeled column density profiles may then look like a hybrid of the models presented here, transitioning between them at a radius of 5-30~AU.

\section{Summary}
We have reported the first detection of HC$_5$N toward the TW Hya protoplanetary disk, representing the largest cyanopolyyne identified to date in a Class II system. We tentatively detected two individual HC$_5$N transitions in ALMA Band 3 with a 5.3$\sigma$ stacked detection, and derived a beam averaged column density $N_T \sim 10^{12} ~cm^{-2}$ with assumed excitation temperatures between 20-50~K. Although our observations do not resolve any radial or vertical structure, known HC$_5$N chemistry in molecular clouds indicates that HC$_5$N should arise in a region where CN and small hydrocarbons coexist. A simple chemical model supports this interpretation, predicting column densities consistent with the observed values and requiring elevated C/O ratios (consistent with previous studies of TW Hya) for HC$_5$N to reach detectable levels. Together, these results extend the known carbon chain inventory of Class II disks and provide new constraints on nitrogen bearing chemistry in planet forming environments.

\section{Acknowledgments}
The National Radio Astronomy Observatory is a facility of the National Science Foundation operated under cooperative agreement by Associated Universities, Inc.  This paper makes use of the following ALMA data: ADS/JAO.ALMA $\#$2019.1.00769.S. ALMA is a partnership of ESO (representing its member states), NSF (USA) and NINS (Japan), together with NRC (Canada) and NSC and ASIAA (Taiwan), in cooperation with the Republic of Chile. The Joint ALMA Observatory is operated by ESO, AUI/NRAO and NAOJ.

L.I.C. also acknowledges support from the Research Corporation for Science Advancement Cottrell Scholarship Award 28249, the David and Lucille Packard Foundation, and NSF AAG Awards 2205698 and 2407547.

Y.A. acknowledges support from JSPS KAKENHI Grant Number 24K00674.

Support for C.J.L. was provided by NASA through the NASA Hubble Fellowship grant No. HST-HF2-51535.001-A awarded by the Space Telescope Science Institute, which is operated by the Association of Universities for Research in Astronomy, Inc., for NASA, under contract NAS5-26555.

A.D. acknowledges support from the NASA FINESST Grant no. 80NSSC24K1470 and the ACM SIGHPC Computational and Data Science Fellowship.

V.V.G. acknowledge support from the ANID -- Millennium Science Initiative Program -- Center Code NCN2024\_001, from FONDECYT Regular 1221352, and ANID CATA-BASAL project FB210003.

C.W.~acknowledges financial support from the Science and Technology Facilities Council and UK Research and Innovation (grant numbers ST/X001016/1 and MR/Z00019X/1).

This work made use of the ChatGPT large language model (OpenAI) for limited assistance with text editing and code formatting.

\software{Astropy \citep{Astropy_2013}, CASA \citep{McMullin_2007}, GoFish \citep{Teague_2019}, Matplotlib \citep{Hunter_2007}, NumPy \citep{Jones_2001}, VISIBLE \citep{Loomis_2018b}, ChatGPT 5.1 \citep{chatgpt_openai}}

\bibliography{refs}{}
\bibliographystyle{aasjournal}

\end{document}